  \documentclass[prl,aps,twocolumn,showpacs]{revtex4}
                                                               \usepackage[dvips]{graphicx}
\usepackage{dcolumn}
                                                                           
\newcommand{\be}{\begin{equation}}
\newcommand{\ee}{\end{equation}}
\newcommand{\bea}{\begin{eqnarray}}
\newcommand{\eea}{\end{eqnarray}}
\newcommand{\nn}{\nonumber}

\begin{document}
\topmargin=-20mm

\title{ Acoustic Cyclotron Resonance and Giant High Frequency
Magnetoacoustic Oscillations in Metals with  Locally Flattened Fermi Surface }
 
\author{Nataliya A. Zimbovskaya and Joseph L. Birman}

\affiliation{Department of Physics, The City College of CUNY, New York, NY, 10031, USA} 
 
\date{\today} 

\begin{abstract}
 We consider the effect of local flattening on the Fermi surface (FS) of a
metal upon geometric oscillations of the velocity and attenuation of
ultrasonic waves in the neighborhood of the acoustic cyclotron resonance.
It is shown that such peculiarities of the local geometry of the FS can
lead to a significant enhancement of both cyclotron resonance and
geometric oscillations. Characteristic features of the coupling of
ultrasound to shortwave cyclotron waves arising due to the local
flattening of the FS are analyzed. 
              \end{abstract}

\pacs{71.18.+y; 72.15.Gd; 72.15.-v }

\maketitle

\section {1. \ Introduction}

Fermi surfaces of  most of  metals are very complex in shape and 
this can essentially influence observables. Those phenomena which are 
determined by the main geometric characteristics of the FS's, i.e. their 
connectivity, are well studied. However those effects which appear due to 
the local geometry of the FS's such as points of  flattening or 
parabolic points have not been investigated in detail at present. 
Meanwhile these local anomalies of the curvature of the FS can noticeably 
affect the electron response of the metal to an external perturbation. The 
change in the response occurs under the nonlocal regime of propagation of 
the disturbance when the mean free path of electrons $ l $ is large 
compared to the wavelength of the disturbance $ \lambda. $ The reason is 
that in this nonlocal regime only those electrons whose motion is somehow
consistent with the propagating perturbation can strongly absorb its
energy. These ''efficient'' electrons are concentrated on small
''effective'' segments of the FS.

When the FS includes points of zero curvature it leads to an enhancement of
the contribution from the neighborhood of these points to the electron
density of states (DOS) on the FS. Usually this enhanced contribution is
small compared to the main term of the DOS which originates from all the
remaining parts of the FS. Therefore it cannot produce noticeable changes
in the response of the metal under the local regime of propagation of the
disturbance $ (l \ll \lambda) $ when all segments of the FS contribute to
the response functions essentially equally. However the contribution to the
DOS from the vicinities of the points of zero curvature can be congruent to
the contribution of a small ''effective'' segment of the FS.  In other
words when the curvature of the FS becomes zero at some points on an
''effective'' part of the FS it can give a noticeable enhancement of
efficient electrons and, in consequence, a noticeable change in the
response of the metal to the disturbance.

The influence of locally flattened or nearly cylindrical segments of the
FS on the attenuation rate and the velocity shift of ultrasonic waves
propagating in a metal, as well as on its surface impedance was analyzed
before (see e.g. references \cite{1,2,3,4}). Some results of this theoretical analysis
were confirmed in the experiments concerning the attenuation of ultrasonic
waves in metals \cite{5,6}. Here we analyze the effect of the local flattening
of the FS on the high frequency magnetoacoustic oscillations. 

 It is known that the absorption coefficient and the velocity of sound
propagating in a metal at right angles to the applied magnetic field $ \bf
B $ in the region of moderately strong magnetic fields for which the
inequalities $ \Omega \tau \gg 1 $ and $ q R \gg 1 $ are satisfied
simultaneously $ (2 R $ is the characteristic diameter of the cyclotron
orbit, and $ q $ is the wave vector of the acoustic wave) oscillate as a
result of variation of the magnetic field. These magnetoacoustic
oscillations, which also are known as geometrical oscillations, are
generated as a result of periodic reproduction of most favorable
conditions for the ''resonance'' absorption of the acoustic wave energy by
electrons moving along the wave front. The oscillations appear due to the
commensurability of cyclotron orbits of the electrons with the wavelength
of the sound wave. Their period is determined by the extremal diameter $ 2
R_{\small \mbox{ex}} $ of the FS of the metal. The geometric oscillations
exist in both low $ (\omega \tau < 1) $ and high $ (\omega \tau > 1) $
grequency ranges $ (\omega $ is the frequency of the sound wave; $ \tau $
is the relaxation time). At high frequencies the magnetoacoustic
oscillations may be superimposed on the acoustic cyclotron resonance.
The main contribution to the oscillating corrections to the attenuation
and the velocity shift originates from the vicinities of so called
stationary points of the cyclotron orbit of the extremal diameter where
an electron moves in parallel to the wave front (figire 1). This leads to a
conjecture that the local geometry of the FS near these stationary points
will strongly affect the geometric oscillations.

\begin{figure}[t]
\begin{center}
\includegraphics[width=6.0cm,height=5cm]{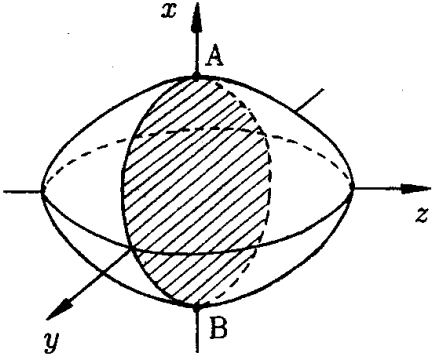}
\caption{The axially symmetric lens corresponding to the energy-momentum relation (16). The lens is flattened at the points $ A(p_2, 0, 0)$ and $ B(-p_2,0,0).$ These points correspond to the stationary points of the cyclotron orbit of the extremal diameter when $ {\bf u}_{q \omega}||{\bf q} || y ,\ {\bf B}|| z.$ }   
\label{rateI}
\end{center}
\end{figure}

It follows from the theory of the geometric oscillations expounded in references \cite{7,8,9} that the amplitude of oscillations in a simple metal with a
closed FS has an order of magnitude smaller by a factor of $ 1 / \sqrt {q R_{\small \mbox{ex}}}$ than smooth components of the absorption coefficient and the velocity of sound. However, in the presence of
certain peculiarities in the geometry of the FS, the number of electrons
effectively participating in the absorption can increase significantly,
leading to an enhancement of the oscillations.  It was proved, for
example in \cite{10,11}, that a sharp increase in the amplitude of geometric
oscillations must take place in a conductor with the Fermi surface in the
form of a slightly corrugated cylinder when the magnetic field is
directed along the cylinder axis.  We show below that when the FS of a
metal is flattened near the points corresponding to stationary points of
the cyclotron orbit of extremal diameter this also leads to a significant
enhancement of the geometric oscillations.

\section{2.\  Acoustoelectronic kinetic coefficients}

Let us consider a longitudinal acoustic wave
propagating in a metal along the $ y $-axis of the coordinate system whose
$ z$-axis is directed along the magnetic field $ \bf B $ and coincides
with a high-order symmetry axis of the crystal. Assume that the elastic
displacement of the lattice $ {\bf u (r,} t)$ is proportional to $ \exp (i
q y - i \omega t).$

Proceeding from basic concepts in the theory of propagation of ultrasound in metals \cite{3,12}, we can write the equation for the amplitude of elastic
displacement $ u_{q \omega} $ of the lattice $ [u (r,t) = u_{q \omega}
\exp (i q y - i \omega t)]$ as: 
            \begin{equation}
- \omega^2 \rho_m u_{q \omega} = - q^2 \rho_m s^2 u_{q \omega} + F_{q
\omega}
                          \end{equation}
 where $ \rho_m $ is the matter density of the lattice. 

 The force exerted by electrons on the lattice contains the contribution originating from their interaction with the electromagnetic field accompanying the sound wave and the deformation contribution. Correspondingly, the magnitude of this force $ F_{q \omega }$ can be written as follows:
      \be 
F_{q \omega} = i q \left (\gamma_\alpha - \frac{i N e}{q} \delta_{\alpha
y} \right ) E_{q \omega}^{'\alpha} + i \omega q^2 \beta u_{q \omega}   
                        \ee
 where 
 $$\displaystyle
{\bf E}_{q \omega}' = {\bf E}_{q \omega} + \frac{i \omega}{c}
\bigg[{\bf u}_{q \omega} \times {\bf B} \bigg ] + \frac{m}{e} \omega^2 {\bf u}_{q \omega};$$
  $ {\bf E}_{q \omega} $ is the amplitude of the  electric field accompanying the wave.

The amplitude $ {\bf E}_{q \omega} $ satisfies the Maxwell equations
      \be 
\bigg [{\bf q \times [q \times E}_{q \omega} ] \bigg ]
= \frac{4 \pi i \omega {\bf J}_{q \omega}}{ c^2}.
                        \ee
 This expression contains the amplitude of the total density of current ${\bf J}_{q \omega} $ induced by the passage of an acoustic wave. The  components of ${\bf J}_{q \omega} $ are given by
      \be 
J_{q \omega}^\alpha = \sigma_{\alpha \beta} E_{q \omega}^{' \beta} +   \omega q \left (\overline \gamma_\alpha - \frac{i N e}{q} \delta_{\alpha y} \right ) u_{q \omega }.
                        \ee
 The electron kinetic coefficients $ \beta $ and $ \sigma $ have the form  
      \bea 
\beta = \frac{i}{2 \pi^2 \hbar^3} \int d p_z m_\perp
\sum \limits_n
\frac{U_{-n} (p_z, - q) U_n (p_z, q)}{\omega + i/\tau - n  \Omega} \\ \nn \\
\sigma_{\alpha \beta} = \frac{i e^2}{2 \pi^2 \hbar^3} \int d p_z m_\perp
\sum \limits_n
\frac{v_{-n}^\alpha (p_z, - q) v_n^\beta (p_z, q)}{\omega + i/\tau - n    
\Omega}
                        \eea
 where $ m_\perp $ is the cyclotron mass and $ U_n (p_z, q) $ is the Fourier transform in the expansion in an azimuthal angle $ \psi $ specifying the position of an electron on the cyclotron orbit:
      \be 
U_n (p_z, q) = \frac{1}{2 \pi} \int _0^{2 \pi} U_n (p_z, \psi, q) \exp (i n \psi) d \psi
                        \ee
 where
      \bea
U_n (p_z, \psi, q)& =& U(p_z, \psi)   
\exp \left [i n \psi - \frac{i q}{\Omega} \int _0^\psi v_y (p_z, \psi') d \psi' \right ]
      \nn \\ \nn \\
U_n (p_z, \psi) &=& \Lambda_{yy} (p_z, \psi) -
 \frac{\big<\Lambda_{yy}\big> - N}{g}.
                        \eea
 Here $ \Lambda_{yy}(p_z, \psi) $ and $ v_y(p_z, \psi) $ are the
corresponding components of the deformation potential tensor and the electron velocity, $ N $ is the electron concentration, the symbol $\big <...\big > $ denotes the averaging over the FS, $ g $ is the density of states on the FS.
        
The Fourier transforms of the electron velocity components in the expansion in the angle $ \psi $ are determined by relations similar to (7)
      \bea
v_n^\alpha (p_z, q) &=& \frac{1}{2 \pi} \int_0^{2 \pi} v_\alpha (p_z, \psi) 
   \nn \\ \nn \\  &\times&
\exp \left [i n \psi - \frac{i q}{\Omega} \int_0^\psi \!\! v_y (p_z,   \psi') d \psi' \right ] d \psi .
                        \eea
 For  a multiply connected FS, the integration with respect to
$ p_z $ in (5) must be supplemented with summation over all cavities of the FS. In this case, the values of $ U_n (p_z, q)$ are calculated separately for each cavity.

In  equations (2), (4) the term $ \displaystyle{({i N e}/{q})}
\delta_{\alpha y} $ then has to be replaced by:
      \be
\frac{i e}{q} \sum \limits_k N_k \frac{e_k}{|e|}
                        \ee
 where a summation has to be performed over the cavities of the Fermi surface; $ N_k $ is the concentration of charge carriers for $ k$-th cavity; $ e_k $ is their charge. When a considered metal has equal number of electrons and holes, the term (10) is equal to zero and the corresponding addends in the expressions for $ {\bf F}_{q \omega}$ and $ {\bf J}_{q \omega}$ vanish. We can obtain the expression for the kinetic coefficient $ \gamma_\alpha $ replacing $ U_{-n} (p_z, -q) $ by $ e v_{-n}^\alpha (p_z, -q) $ in Eq.(5). To obtain the expression for $ \overline \gamma_\alpha $ we have to replace $ U_n(p_z, q)$ by $ e
v_n^\alpha (p_z, q).$

To determine the wave vector of the acoustic wave propagating in a metal we have to solve the equation for the amplitude of elastic displacement of the lattice together with the Maxwell equations. As a result we arrive at the formula
      \be
q^2 = \frac{\omega^2}{s^2} - \frac{i \omega q^2}{\rho_m s^2}
\left (\beta^* + \frac{\gamma^* \big(\overline \gamma^* - B c q/ 4 \pi \omega \big)}{\sigma^* - c^2 q^2 / 4 \pi i \omega} \right ).
                        \ee
  Here 
      \bea &&  
\beta^* = \beta - \left [\gamma_y - \frac{i e}{q} \sum \limits_k N_k \frac{e_k}{|e|} \right ]^2 \Big / \sigma_{yy}\, ,
         \nn \\ \nn \\&&
\gamma^* = \gamma_x - \left [\gamma_y - \frac{i e}{q} \sum \limits_k N_k \frac{e_k}{|e|} \right ] \frac{\sigma_{yx}}{\sigma_{yy}}\, ,
                      \nn \\ \nn \\&&
\overline \gamma^* = \overline \gamma_x
- \left [\overline \gamma_y - \frac{i e}{q} \sum \limits_k N_k
\frac{e_k}{|e|} \right ] \frac{\sigma_{yx}}{\sigma_{yy}}\, ,
                   \nn \\ \nn \\ &&
\sigma^* = \sigma_{xx} + \sigma_{yx}^2 / \sigma_{yy}.
                        \eea

For small amplitudes of acoustic waves, the wave vector is described by the expression
      \be 
q = \omega /s + \Delta q.
                        \ee
 The increment $ \Delta q $ linear in $ u_{q \omega} $ which emerges as a result of interaction with electrons, has the form in the case under investigation 
      \be
\Delta q =  \frac{i q^2}{2 \rho_m s}
\left (\beta^* + \frac{\gamma^* \big(\overline \gamma^* - B c q/4 \pi \omega \big )}{\sigma^* - c^2 q^2 / 4 \pi i \omega} \right ).
                        \ee
 The wave vector  $ q $ on the right-hand side of (14) is assumed to be equal to $ \omega /s. $

In the region under investigation where $ d R \gg 1,$ the main contribution to the integral with respect to $ \psi $ in expressions (7), (9) for $ U_n(p_z, q)$ and $ v_n^\alpha (p_z, q) $ comes from the neighborhoods of   stationary points on cyclotron orbits. Accordingly, estimating the integrals by the stationary phase method, we can obtain the following
asymptotic expressions for $ U_{\pm n} (p_z, \pm q): $
      \bea 
U_{\pm n} (p_z, \pm q)& =& \frac{1}{\pi} U_0 (p_z) \exp \left [
\pm i q R(p_z) \pm i \pi \frac{n}{2} \right ] 
        \nn \\ \nn \\ & \times &
\left \{ \cos \left (q R(p_z) - \pi \frac{n}{2} \right ) V (p_z) \right.
         \nn \\ \nn \\ &-& \left.
\sin \left (q R(p_z) - \pi \frac{n}{2} \right ) W (p_z) \right \}
                        \eea
 where $ U_0 (p_z) = U (p_z, \psi_1) = U(p_z, \psi_2), \ 2R $ is the diameter of a cyclotron orbit of electrons   in the direction of propagation of the acoustic wave, $ \psi_1 $ and $ \psi_2 $ are the values of the angle $ \psi $ corresponding to stationary points on the cyclotron orbit, $ \psi_1 - \psi_2 = \pi.$ The form of the functions $ V (p_z) $ and $ W (p_z) $ is determined by singularities of the energy momentum relation for electrons in the vicinity of stationary points.

\section  
{3. \ The model and results}
  
 Let us assume that among cavities of a closed Fermi surface there is a biconvex lens, whose symmetry axis is the $x$-axis of chosen coordinate system.  We write the energy momentum relation for the electrons associated with the lens in the form
  \be 
E({\bf p}) = \frac{p_1^2}{2m_1} \left
(\frac{p^2_ y + p^2_z}{p^2_1} \right)^r +
\frac{p^2_2}{2m_2} \left(\frac{p_x}{p_2}
\right)^2 .
                                        \ee
 The curvature of the FS is given by the formula:
      \be 
K = \frac{1}{v^3} (P R - Q^2)
                                       \ee
 where $ v = \sqrt{v_x^2 + v_y^2 + v_z^2} ;$ also
      \bea &&
P = \left (\frac{\partial v_z}{\partial p_x} + 
\frac{\partial v_x}{\partial p_z} \right )v_x v_z -
\frac{\partial v_z}{\partial p_z} v_x^2 -
\frac{\partial v_x}{\partial p_x} v_z^2; 
       \nn \\ \nn \\ &&
Q = \left (\frac{\partial v_z}{\partial p_x}v_y + 
\frac{\partial v_z}{\partial p_y} v_x \right )v_z -
\frac{\partial v_z}{\partial p_z} v_x v_y -
\frac{\partial v_y}{\partial p_x} v_z^2; 
            \nn \\ \nn \\ &&             
R = \left (\frac{\partial v_z}{\partial p_y} + 
\frac{\partial v_y}{\partial p_z} \right )v_y v_z -
\frac{\partial v_z}{\partial p_z} v_y^2 -
\frac{\partial v_y}{\partial p_y} v_z^2. 
                                        \eea
 Straightforward calculations give us the following result for the curvature of the lens:
      \bea 
K &=& \frac{r}{m_1 v^4} \left (\frac{p_y^2 + p_z^2}{p_1^2} \right )^{r-1} 
   \nn \\ \nn \\ &\times&
 \left [\left (\frac{v_y^2 + v_z^2}{m_2} \right )
\frac{\partial v_x}{\partial p_x} + 
\frac{r (2r - 1)}{m_1}
\left (\frac{p_y^2 +p_z^2}{p_1^2} \right)^{r-1} \! \! v_x^2 \right ]. \nn \\
                                       \eea
 If the parameter $ r $ characterizing the shape of the lens assumes values greater than unity, then the Gaussian curvature of the surface vanishes at the points $({\pm}p_2;0;0)$, which coincide with the vertices of the lens.  Because of the axial symmetry of the lens, the vertices are points where the surface of the lens is flattened. The lens will be flatter near its vertices, the greater the value of $ r.$ The results of
experiments on the cyclotron resonance in a magnetic field applied along a normal to the surface of a metal \cite{13,14} give a basis for the conjecture that such flattened electron lens may be an element of the FS's of cadmium and zinc \cite{15}. 

Electrons associated with the vicinities of the vertices of the lens will strongly participate in the absorption of the energy of the acoustic wave, when both the magnetic field and the acoustic wave vector are perpendicular to the axis of the lens (figure 1). Within the framework of the model (16) functions $ V (p_z) $ and $ W (p_z) $ included into the expression (15) for $ U_{\pm n} (p_z, \pm q) $ corresponding to the lens can be written as follows: 
   \bea 
V (p_z) = \int_0^\infty \cos \bigg [q R (p_z) Q_r (y, p_z) \bigg
]dy 
                 \nn \\ \nn \\          
W (p_z) = \int _0^\infty \sin \bigg [q R (p_z) Q_r (y, p_z) \bigg ]dy
                            \eea
 where
  \be 
Q_r (y, p_z) = \sum \limits_{k=1}^r a_k (p_z) y^{2k}
\left (\frac{m_\perp^2}{m_1 m_2} \right )^k 
                                         \ee
 and $ m_\perp $ is the cyclotron mass for the electrons associated with the lens. All dimensionless coefficients $ a_k (p_z) $ except $ a_r (p_z) $ become zero at $ p_z = 0; $ the latter is of the order of unity at this
point.  Specifically for $ r = 2 $ we have
  \be 
 Q_2 (y, p_z) = \frac{m_\perp^2}{m_1 m_2} a_1 (p_z) y^2 + \left ( \frac{m_\perp^2}{m_1 m_2} \right )^2 a_2 (p_z) y^4
                            \ee
     where
  \be 
a_1 (p_z) = \frac{p_z^2}{p_1^2}, \qquad 
a_2 (p_z) = \frac{1}{2} \left (1 - \frac{4}{3} \frac{p_z^4}{p_1^4} \right).
                           \ee
 The leading term of the asymptotic expansion of the function
$V(p_z)$ in inverse powers of $ qR $ originates from a neighborhood of the central cross section. For $ r = 2 $ it has the form 
  \be 
V(p_z) = \frac{\Gamma(1/4)}{4}\frac{\sqrt{m_1m_2}}
{m_\perp^{\small \mbox{ex}}}\left(\frac{2}{qR_{\small \mbox{ex}}} \right)^{1/4}\cos
\left (\frac{\pi}{8} \right ).
                     \ee
 Here, $\Gamma (x)$ is the gamma function, $ m_\perp^{ex} = m_\perp (0)$, and $R_{ex} = R (0)$. The asymptotic expression for $W(p_z)$ is obtained from equation (24)  by replacing the cosine by a sine of the same argument. 
  
 For an arbitrary value of $ r $, the function $V(p_z)$ in a neighborhood of $p_z = 0$ is described by the asymptotic expression 
    \be 
V(p_z) = \frac{1}{2r}\frac{\Gamma(1/{2r})}{(qR_{ex})
^{1/{2r}}}\sqrt{\frac{m_1m_2}{m_\perp^{ex}(a_r(0))^{1/r}}}\cos(\pi/{4r}).
                    \ee  
 A similar expression can also be written for  $W(p_z)$.

Using the asymptotic expressions (15), (25) for $ U_{\pm n} (p_z, \pm q) $ and similar asymptotics for $ v_{\pm n}^\alpha (p_z, \pm q) $ we arrive at expressions for electron kinetic coefficients. Taking intoaccount that the largest contribution to the integrals over $ p_z $ in the expressions (5), (6) originates from the range of small $ p_z, $ we can replace all smooth functions of $ p_z $ in the integrands by their values at $ p_z = 0.$ For $ q R \gg 1 $ the main contribution to the
asymptotic expression for $ \beta $ is associated with the electrons of the lens:
  \be 
\beta = \frac{i g}{\omega} \frac{\mu}{(q R_{ex})^{1/r}} U_0^2 (0) X(\omega).
                            \ee
 Here $ R_{ex} = R(0);$ a dimensionless contant $ \mu $ equals:
  \be            
 \mu = \frac{a_r^2}{4 \sqrt \pi} \frac{\big<1\big>}{g} \Gamma \left (\frac{r + 1}{2 r} \right ) \Big / \sqrt{\int_0^1 \overline m_\perp (x) d x.}
                            \ee
 We introduce the notation $\big <1\big> $ for the electron DOS on the lens; $ x = p_z /p_m ; \ \overline m_\perp (x) = m_\perp (x) / m_\perp (0). $ The frequency-dependent factor $ X (\omega) $ in equation (26)  has the form
  \be 
X (\omega) = \int_{-1}^1 Y (\omega, x) dx 
                            \ee
 where
  \bea
Y (\omega ,x) &=& - i \pi \frac{\omega}{\Omega} \left \{
\coth \left [\pi \frac{1 - i \omega \tau}{\Omega \tau } \right ] \right. 
     \nn \\ \nn \\    & + & \left.
\cos \left (2 q R + \frac{\pi}{2 r} \right ) \left (
\sinh \left [\pi \frac{1 - i \omega \tau}{\Omega \tau } \right ]
\right )^{-1} \right \} . \nn \\
                            \eea

We obtain asymptotic expressions for remaining electroacoustic kinetic coefficients in similar way. Specifically we have
  \be 
\gamma_y = \frac{ie}{q} \sum \limits_k N_k \frac{e_k}{|e|} + \frac{ie g}{q} \frac{\mu}{(q R_{ex})^{1/r}} U_0 (0) X (\omega) 
             \ee \be
\sigma_{yy} = - \frac{i e^2}{q^2} \omega g \left (1 -
\frac{\mu}{(qR_{ex})^{1/r}} X (\omega) \right ).
                            \ee
  The oscillating terms in equations (30), (31) are mainly determined by the contributions from the flattened electron lens. Contributions from  remaining (nonflattened) cavities of the FS are proportional to the small factor $ 1/ qR$ and we can omit them.

In the  high-frequency range $ (\omega \tau \gg 1) $ the function $ Y (\omega, x) $ has singularities at frequencies $ \omega $ which are equal to the multiple cyclotron frequency $ \Omega. $ These singularities arise due to the acoustic cyclotron resonance which was analyzed in references \cite{16,17}.  The second term in equation (29) also contains the factor $ \cos (2qR + \pi /2r)$ describing geometric oscillations.

The main contribution to the integral (28) is from the region of small $ x $ where the cyclotron frequency is close to its extremum value $ \Omega_{ex}. $ In this region which corresponds to the vicinity of the central cross-section of the lens, we can use the following approximation:
  \be 
\Omega (x) = \Omega_{ex} (1 + \eta^2 x^2)
                            \ee
 where
  \be 
\eta^2 = \frac{1}{\sqrt \pi} \frac{\Gamma \big ((r + 1)/2r 
\big )}{\Gamma (1 + 1/2r)} \int_0^1 \frac{dz}{z^2} \left
(\frac{1}{\sqrt{1 - z^2}} - \frac{1}{\sqrt{1 - z^{2r}}}\right ).
                            \ee
 When $ r = 1 $ and the lens is ellipsoidal in shape this parameter $ \eta^2 $ becomes  zero. In this case the cyclotron frequency is independent of $ p_z. $ For a flattened lens $ (r \ge 2)$ this parameter takes nonzero values which may be of the order of unity.
 
\subsection
{3.1.\ Case A: moderately flattened FS}

The asymptotic expression for the function $ X (\omega)$ near the cyclotron resonance depends on the ratio of the parameters $ 2 q R_{ex}$ and $ (\omega \tau)^{r/2}.$ Under considered conditions both parameters are large compared to unity. Suppose that $ 2 q R_{ex} \gg (\omega \tau)^{r/2}.$ Under conditions of the acoustic cyclotron resonance in typical metals the parameter $ q R_{ex} \sim v_F/s \sim 10^3 \ (v_F $ is
the Fermi velocity for the electrons associated with the lens). For $ \Omega \tau \sim 10^2 $ this inequality can be satisfied when the lens is moderately flattened $ (1 < r < 2).$ The asymptotic expression for the function $ X (\omega) $ near the cyclotron resonance can be written as follows: 
  \be 
X (\omega) = \frac{\pi}{\eta} \frac{1}{\aleph} \left [1 +
\frac{(-1)^nb}{(q R_{ex})^{1/2l}}
\frac{\cos (2q R_{ex} + \pi/4r)}{\aleph} \right ]
                            \ee
 where 
  $$ \displaystyle{b = \frac{2 \eta}{\pi} \Gamma (1 + 1/2r)}; \qquad
 \aleph = \sqrt{1 - \frac{\omega}{n \Omega_{ex}} - \frac{i}{\omega \tau}}.
  $$
  The principal term in the obtained expression for $ X (\omega ) $ is its first term. The second term in equation (34) which describes the geometric  oscillations is significantly smaller in magnitude.

When $ 2 qR_{ex} \gg (\omega \tau)^{r/2} $ the dynamical correction $    \Delta q $ near the acoustic cyclotron resonance remains small compared to the main approximation of the ultrasound wave vector $ \omega /s .$ For  longitudinal waves this correction is mainly determined by the deformation
interaction of the sound wave with the electrons. The resonance
contribution to the correction $ \Delta q $ from the electrons associated with the neighborhood of the central cross-section of the lens (16) equals
  \bea 
\Delta q &=& \gamma_0 \frac{1}{(qR_{ex})^{1/r}} \frac{q R_{ex}}{n}  \frac{1}{\aleph} 
   \nn \\ \nn \\ & \times &
\left [1 + \frac{b \cos(2q R_{ex} + \pi n + \pi /
4r)}{\aleph( q R_{ex})^{1/2r} } \right].
                            \eea
 Here 
  $$ \gamma_0 = \displaystyle{\frac{\pi N q \omega m_\perp (0) \mu}{2 \eta \rho_m s^2 p_2}} U_0^2 (0) 
    $$
  is the quantity of the dimensions and order of the attenuation rate for high frequency ultrasound waves in the
absence of the external magnetic field.

The real and imaginary parts of the correction $ \Delta q $ determine  the resonance contributions from the electrons associated with the lens to the velocity shift $ \Delta s/s $ and the attenuation rate $ \Gamma $ of the ultrasound wave:
  \be
\frac{\Delta q}{q} = \frac{\Delta s}{s}  + \frac{i \Gamma}{2 q}.
                      \ee
 For $ r = 1$ the result (35) for the attenuation rate coincides with the corresponding result of reference \cite{17}, which is obtained under assumption that the FS of a metal everywhere has a finite and nonzero curvature. When $ l = 1 $ the magnitude of the resonance feature in the attenuation rate is of the order of $ \gamma_0 \sqrt{\omega \tau} /n.$ In this case the magnitude of the geometrical oscillations is smaller by a factor $ \sqrt{\omega \tau/qR_{ex}}$ than the magnitude of the resonance feature connected with the cyclotron resonance.

When $ r > 1$ the effective strip on the FS passes through the
flattened segments near the vertices of the lens. It gives the
amplification of the acoustic cyclotron resonance. The dependence of the attenuation rate of the ultrasound on the magnetic field near the cyclotron resonance is shown in the figure 2. The FS is assumed  to be moderately flattened. The resonance contribution to the ultrasonic  absorption coefficient increases $ (qR_{ex})^{(r-1)/r} $ times. This
amplification arises due to the increase in the number of electrons participating in the resonance absorption of the energy of ultrasound wave.

\begin{figure}[t]
\begin{center}
\includegraphics[width=6.0cm,height=8cm]{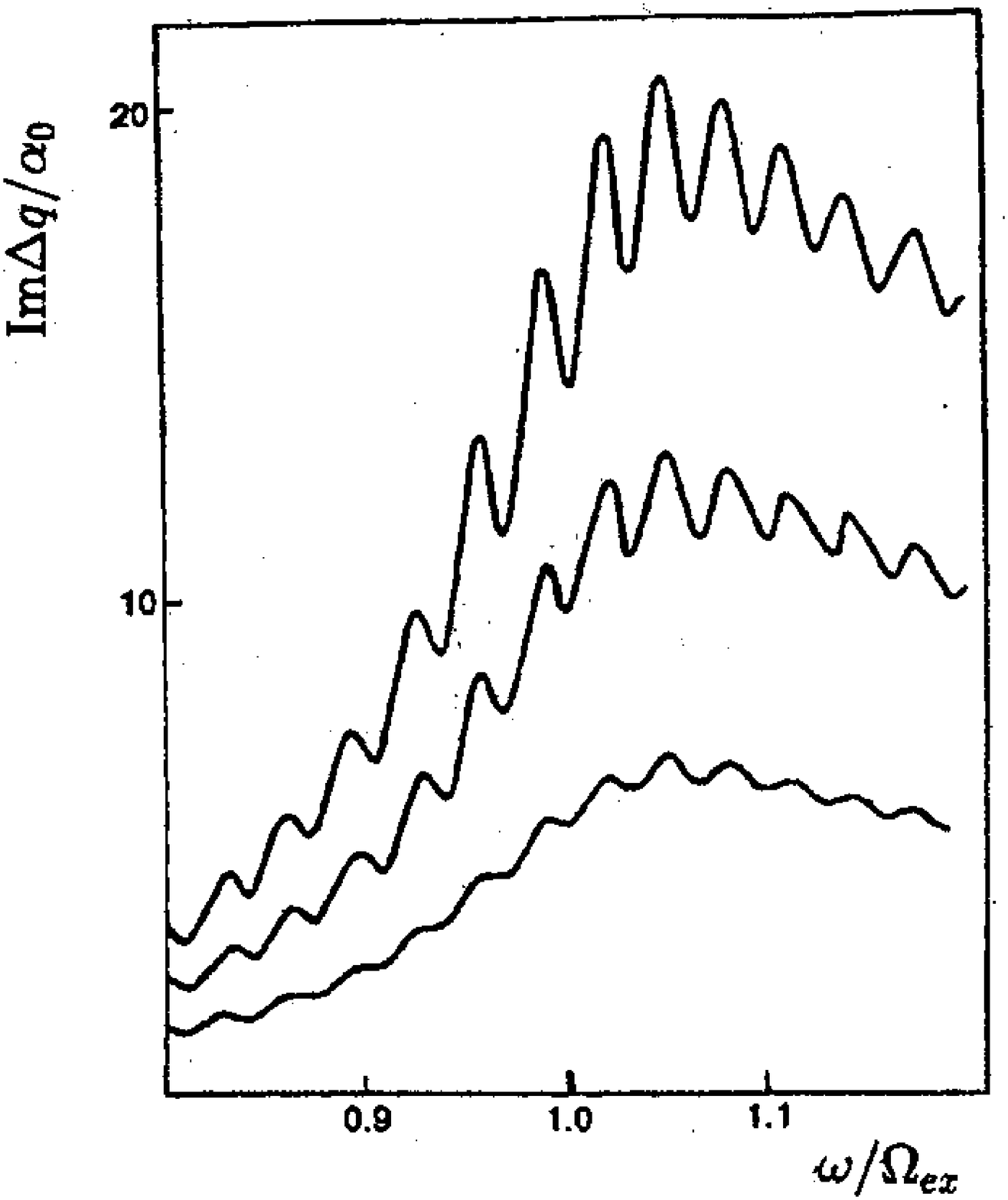}
\caption{Attenuation of longitudinal ultrasound waves versus $ \omega /\Omega_{ex}$ in the vicinity of the cyclotron resonance for a moderate flattening  of the Fermi surface near vertices of the electron lens. Curves are plotted for $ \omega \tau = 10, \ q R_{ex} = 100,\ r = 1.25  $ (curve 1), $ r = 1.5 $ (curve 2), $ r = 1.75 $ (curve 3).
}
\label{rateI}
\end{center}
\end{figure}

This increase in the number of efficient electrons also leads to the amplification of the geometric oscillations. The corresponding term in equation (35) is $ (qR_{ex})^{(r-1)/2} $ times larger in magnitude than a similar term in the expression for $ \Delta q $ in a simple metal. When the flattening of the FS becomes stronger, the magnitude of the geometric
oscillations grows faster than the magnitude of the peak corresponding to the acoustic cyclotron resonance. The larger $ r $ the relatively larger is the contribution from the term associated with the geometrical oscillations (the second term in the expression (35)) to the correction $\Delta q.$
 
\subsection
{3.2.\ Case B: strongly flattened FS}

We can use the expression (35) to describe the resonance part of the dynamic correction $ \Delta q$ only for moderate flattening of the electron lens and moderately large $ \omega \tau. $ When the flattening of the lens near its vertices is strong, the quantity $ (\omega \tau)^{r/2} $ exceeds the parameter $ 2q R_{ex}.$ Under the conditions of acoustic
cyclotron resonance in typical metals the inequality $ 2q R_{ex } \ll (\omega \tau)^{r/2} $ can be satisfied for $ r > 3.$

In this case we have to use a new asymptotic expression for the function $ X (\omega).$ This new asymptotic can be written as follows:
  \be 
X (\omega) = \frac{\pi}{\eta} \frac{1}{\aleph} \left [1 + (-1)^n \cos \left (2q R_{ex} + \frac{\pi}{2r} \right ) \right ].
                            \ee
 Both terms in this expression (37) are of the same order in magnitude. It critically changes the magnetic field dependence of the function $ X (\omega ) $ near the resonance $ (\omega \approx n \Omega_{ex}).$ When the asymptotic (37) is applicable, the factor $ X (\omega) /(qR_{ex})^{1/r}$
in the expressions for the kinetic coefficients is not small compared to unity. In this connection the contribution to the dynamic correction $ \Delta q $ arising due to the interaction with the electromagnetic field  accompanying the sound wave becomes significant.

The effects originating from coupling of electromagnetic and ultrasound waves are well known. Specifically it is shown that the ultrasound wave propagating perpendicularly to the external magnetic field can couple to shortwave cyclotron waves (see references \cite{18,19,20}). In our geometry longitudinal
ultrasound waves couple to longitudinal cyclotron wave whose dispersion relation is determined by the equation $ \sigma_{yy} = 0. $ The dispersion curve of this mode near the frequency $ n \Omega_{ex}$ can be written in the form
  \be 
\omega_1 = n \Omega_{ex} \left [1 - \frac{1}{(qR_{ex})^{2/r}} f^2 (q) \right ].
                            \ee
 Here $ f (q) $ is an oscillating function:
  \be 
f (q) = \frac{2 \pi \mu}{\eta} \cos^2 \left [q R_{ex} + \frac{\pi n}{2} + \frac{\pi}{4 r} \right ].
                            \ee
 This cyclotron mode can propagate in a metal under the condition $ 2qR_{ex} \ll (\omega \tau)^{r/2}. $ The shape of the dispersion curve of the considered cyclotron wave depends on the local geometry of the Fermi surface. Longitudinal cyclotron waves similar to the mode described by equation (38) can propagate in a metal with a spherical FS under the condition $ qR_{ex} < \omega \tau. $ Their dispersion relation has the form (see reference \cite{18}) 
  \be 
\omega_1 = n \Omega \big (1 + 1 /2 q R \big ).
                            \ee
 The difference in the expressions (38) and (40) describing the
dispersion curves of the longitudinal cyclotron waves are completely caused by the local flattening of the considered FS.

For a very strong flattening of the vicinities of the vertices of electron lens $ (2 q R_{ex} \ll (\omega \tau)^{r/2}) $ we can write the following expression for the resonance contribution to the dynamic   correction $ \Delta q:$
  \be 
\Delta q = \gamma_0 \frac{q R_{ex}}{n} \frac{f^2 (q)}{(q R_{ex})^{2/r}} \frac{\omega}{\omega_1 - \omega - i/\tau}.
                            \ee
 Here $ \omega_1 $ is the frequency of the longitudinal cyclotron wave described by the formula (41).  The frequency $ \omega_1 $ corresponds to the resonance rather than the cyclotron frequency $ \Omega_{ex}. $ The shift of the peak of the acoustic cyclotron resonance caused by the coupling of the ultrasound to the shortwave cyclotron wave was studied for
the spherical and ellipsoidal FS's.  When the effective segments of the FS are locally flattened this shift is more pronounced and more available for experimental observations. 

Besides the cyclotron mode described by equation (41), Fermi-liquid cyclotron waves can propagate in metals. Coupling to these Fermi-liquid modes can change the resonance contribution to the dynamical correction $ \Delta q $ near the acoustic cyclotron resonance. However, it is shown in reference \cite{21}
that these changes are not very significant because the coupling of the ultrasound to these Fermi-liquid modes is more weaker than to the mode analyzed above. It gives reasons to neglect Fermi-liquid effects in the present consideration.

The factor $ f^2 (q) $ in the expression (41) describes the geometric oscillations which are superimposed on the peak corresponding to the acoustic cyclotron resonance. The amplitude of these geometric oscillations sharply increases near the resonance. In order of magnitude it is determined by the height of the resonance peak. Thus the geometric oscillations of the ultrasonic absorption coefficient in metals with strongly flattened FS's can reach values of the order of the smooth part of the attenuation rate. The geometric oscillations may become giant near the acoustic cyclotron resonance. Figure 3 illustrates this conclusion.

\begin{figure}[t]
\begin{center}
\includegraphics[width=6.0cm,height=8cm]{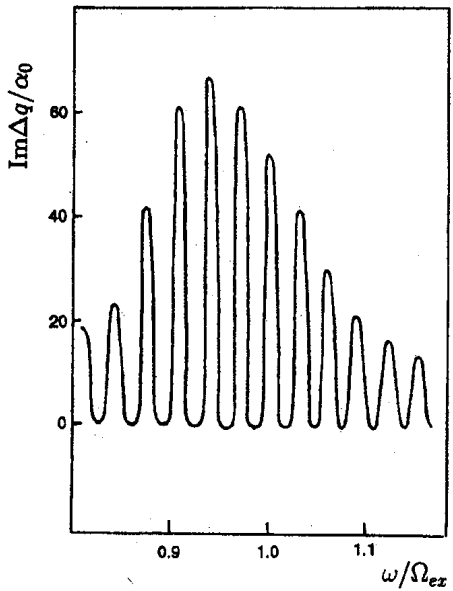}
\caption{Giant geometrical oscillations of the attenuation rate of 
longitudinal ulterasound waves in the vicinity of the cyclotron resonance 
for strong flattening of the Fermi surface near vertices of the electron 
lens $(r = 4).$ Curve is plotted for $ \omega \tau = 10, \; q R_{ex} = 
100, \ \mu = 0.1. $ Shift of the resonance occurs due to the coupling 
of the ultrasound to the cyclotron wave.
}
\label{rateI}
\end{center}
\end{figure}
Since the amplification of the geometric resonances in the velocity and absorption of ultrasound is due to the local geometric characteristics of the FS it can be observed only for a definite choice of the direction of the magnetic field with respect to the symmetry axes of the crystal lattice. When the magnetic field is tilted away from the direction for which the point of flattening of the FS falls on its section corresponding
to the cyclotron orbit of the electrons participating effectively in the formation of the oscillations, the influence of this point vanishes and the amplitude of the oscillations descreases. Therefore, the amplification of geometric oscillations, just as a number of other effects which result
from local geometric features of the FS of a metal \cite{1,2,3,4}, should exhibit a pronounced dependence on the direction of the external magnetic field. Specifically, for the model FS (16) considered here, the amplitude of the
geometric oscillations of the velocity and attenuation rate of the sound will depend on the angle $ \varphi $ between the external magnetic field and a plane perpendicular to the axis of the lens. The range of variation of the amplitude of the oscillations on increasing $ \varphi $ is determined by the degree of flattening of the lens near its vertex. 

\section
{4.\ summary}

In  summary, the results of study of electron energy spectra of metals aswell as the experimental results of references \cite{5,6,13,14} give the basis for the assumption that the FS's of certain metals  (e.g. cadmium and zink) can include locally flat segments. These feature of the local geometry of the FS's can be enhanced by applying an agent that changes the shape of
the constant energy surfaces, e.g., external pressure. Also the local flattening of the FS can be enhanced in ultra-thin films of metals. We showed that the local flattening of the FS of a metal can give rise to an essential amplification of both acoustic cyclotron resonance and geometric oscillations of the attenuation rate and the velocity shift of the ultrasound wave. We predict that the amplification has to be particularly strong for a strong local flattening of the FS.  When the flattening of the FS at the stationary points is strong enough (within the
framework of our model (16) this corresponds to $ r > 3$) the magnitude of the geometric oscillations of the attenuation rate of the ultrasonic waves near the acoustic cyclotron resonance can reach values of the order of smooth part of the absorption coefficient. We also predict that for a strongly flattened FS the shift in the resonance frequency of the cyclotron resonance which occurs due to the coupling of the ultrasound to the shortwave cyclotron wave has to be more pronounced than in typical metals and can be observed in experiments. The probability of the observation of the effect of enhancement of the geometric oscillations increases due to the anisotropy of this effect.  The magnitude of the oscillations depends on the orientation of the magnetic field. The appearance of such a dependence can give an experimental evidence of the effect and also additional information on the geometry of the FS of some
metals.

\section{ Acknowledgments}

We thank Dr. G.M. Zimbovsky for help with the manuscript. Support from a PSC-CUNY FRAP "In -- Service" Award is acknowledged.

\end{document}